\documentclass[aps,prd,reprint,twocolumn,superscriptaddress,longbibliography,nofootinbib,floatfix,showpacs]{revtex4-2}

\usepackage{graphicx}
\usepackage{amsmath,amssymb,amsfonts}
\usepackage{hyperref}
\usepackage{slashed}
\usepackage{bm}
\usepackage[usenames]{color}
\usepackage[T1]{fontenc} \usepackage[latin1]{inputenc}

\usepackage{listings}
\usepackage[scaled=0.85]{inconsolata}  

\newcommand{\be}{\begin{equation}}
\newcommand{\ee}{\end{equation}}
\newcommand{\ud}{\mathrm{d}}

\newcommand{\LCm}{{\scriptscriptstyle -}} 
\newcommand{\LCp}{{\scriptscriptstyle +}}
\newcommand{\LCpm}{{\scriptscriptstyle\pm}}
\newcommand{\LCperp}{{\scriptscriptstyle \perp}}
\newcommand{\LCpara}{{\scriptscriptstyle \parallel}}

\begin{document}

\title{Solving the Dirac equation on a GPU for strong-field processes in multidimensional background fields}
\author{Greger Torgrimsson}
\email{greger.torgrimsson@umu.se}
\affiliation{Department of Physics, Ume{\aa} University, SE-901 87 Ume{\aa}, Sweden}

\begin{abstract}

In this paper, we show how to solve the Dirac equation, $(i\gamma^\mu[\partial_\mu+ieA_\mu(t,{\bf x})]-m)\psi=0$, on a GPU. This is orders of magnitude faster than solving it on CPU and allows us to consider background fields, $A_\mu(t,{\bf x})$, that depend on $2+1$ or even $3+1$ coordinates. Our approach is conveniently implemented using the computational library JAX. We show how to obtain the probabilities of Schwinger and nonlinear Breit-Wheeler pair production from these solutions using a scattered-wave-function approach and compare the results with the worldline-instanton approximations.   

\end{abstract}

\maketitle

\section{Introduction}

Processes in strong background fields can be studied with the Furry-picture expansion, where the coupling to the background field, $eF_{\mu\nu}$, is treated exactly, while radiative corrections are treated perturbatively. By rescaling the background as $e F_{\mu\nu}\to F_{\mu\nu}$, the probabilities are expanded as ($\alpha=e^2/4\pi$)
\be
P=\sum_n \alpha^n P_n(F) \;.
\ee
For spontaneous/Sauter-Schwinger pair production ($\to e^\LCp e^\LCm$), the expansion starts at $n=0$. For nonlinear Compton scattering ($e^\LCm\to e^\LCm\gamma$) and nonlinear Breit-Wheeler pair production ($\gamma\to e^\LCp e^\LCm$), the expansions start at $n=1$. For higher orders, see~\cite{Fedotov:2022ely}. The functions $P_n(F)$ can be obtained by solving the dressed Dirac equation
\be\label{DiracEq}
(i\slashed{D}-1)\psi=0 \;,
\ee
where $D_\mu=\partial_\mu+iA_\mu$ and we use units where $c=\hbar=m_e=1$. It has been numerically challenging to solve~\eqref{DiracEq} for multidimensional fields, so the state-of-the-art has for some time been Schwinger pair production for fields that only depend on time and one spatial coordinate~\cite{Hebenstreit:2011wk,Jiang:2012mfn,Jiang:2013wha,Wollert:2015kra,Kohlfurst:2015zxi,Kohlfurst:2015niu,Aleksandrov:2016lxd,Aleksandrov:2017mtq,Kohlfurst:2017hbd,Kohlfurst:2017git,Lv:2018wpn,Peng:2018hmj,Ababekri:2019dkl,Kohlfurst:2019mag,Aleksandrov:2019ddt,Ababekri:2019qiw,Li:2021vjf,Mohamedsedik:2021pzb,Li:2021wag,Kohlfurst:2021skr,Jiang:2023hbo}.
In this paper we show how to combine the scattered-wave-function (SWF) approach in~\cite{Torgrimsson:2025pao} with modern and powerful GPU tools to study both $2+1$ and $3+1$ dimensional fields. We also extend SWF to nonlinear Breit-Wheeler. We compare these fully numerical results with the weak-field approximations obtained using the open-worldline-instanton methods from~\cite{DegliEsposti:2021its,DegliEsposti:2022yqw,DegliEsposti:2023qqu,DegliEsposti:2023fbv,DegliEsposti:2024upq}.

\section{Scattered wave functions}

To solve the Dirac equation numerically we use the SWF approach~\cite{Torgrimsson:2025pao}. We seek solutions to~\eqref{DiracEq} with plane-wave initial conditions in the asymptotic past or asymptotic future, 
\be
\begin{split}
\lim_{t\to-\infty}U_{\rm in}(s{\bf p}x)&=u_s({\bf p})e^{-ipx}\\
\lim_{t\to-\infty}V_{\rm in}(s{\bf p}x)&=v_s({\bf p})e^{ipx}\\
\lim_{t\to+\infty}U_{\rm out}(s{\bf p}x)&=u_s({\bf p})e^{-ipx}\\
\lim_{t\to+\infty}V_{\rm out}(s{\bf p}x)&=v_s({\bf p})e^{ipx} \;.
\end{split}
\ee
We assume that the gauge potential vanishes asymptotically in all directions, $A_\mu\to0$ as $|x^\mu|\to\infty$, so $U_{\rm out}(t{\bf x})$ is equal to $u_s({\bf p})e^{-ipx}$ not just at $t\to\infty$ but also for finite $t$ as long as $|{\bf x}|$ is sufficiently large, and similar for the other wave functions. 

In~\cite{Torgrimsson:2025pao} we considered cases such as $A_1(t,x)=(E/\omega)\tanh(\omega t)\text{sech}^2(\kappa x)$ for which $A_1(\infty,x)\ne A_1(-\infty,x)\ne0$. However, there is less motivation to consider such cases in $(2+1)$D or $(3+1)$D because the asymptotic space outside the field has a different topology, which affects what type of gauge fields are relevant. In 1D, e.g. $E(t)$ or $E(x)$, the asymptotic space is two disconnected regions, $t<-|R|$ and $t>|R|$, where $R\gg1$. In 2D, e.g. $E(t,x)$, the asymptotic space is an annulus, $t^2+x^2>R$, which is not simply connected. In 3D and 4D, the asymptotic space is simply connected. It is therefore well motivated to consider $A_\mu\to0$ as $|x^\mu|\to\infty$.       

We split the wave function into a background plane wave and scattered wave,
\be
\psi=\psi_{\rm back}+\psi_{\rm scat} \;,
\ee
where
\be
\begin{split}
U_{\rm out}^{\rm back}&=U_{\rm in}^{\rm back}=ue^{-ipx} \\
V_{\rm out}^{\rm back}&=V_{\rm in}^{\rm back}=ve^{ipx} \;.
\end{split}
\ee
The scattered waves are determined by an inhomogenous PDE,
\be\label{scatDirac}
(i\slashed{D}-1)\psi_{\rm scat}=-(i\slashed{D}-1)\psi_{\rm back}=\slashed{A}\psi_{\rm back} \;,
\ee
with initial conditions
\be
\lim_{t\to-\infty}\psi_{\rm in}^{\rm scat}=\lim_{t\to\infty}\psi_{\rm out}^{\rm scat}=0 \;.
\ee
The scattered waves have finite support in ${\bf x}$, which makes this formulation particularly suitable for numerics.

The probability to produce an electron with $m=(s{\bf p})$ and a positron with $n=(r{\bf q})$ is given by~\cite{Torgrimsson:2025pao}
\be\label{Nmn}
\begin{split}
N(m,n)=&\Big|{}_m(U_{\rm back}|V_{\rm scat}^{\rm out})_n\\
&+{}_m(U_{\rm scat}^{\rm out}|U_{\rm back})_i{}_i(U_{\rm back}|V_{\rm scat}^{\rm out})_n\Big|^2 \\
&\Big|{}_m(U_{\rm scat}^{\rm out}|V_{\rm back})_n\\
&+{}_m(U_{\rm scat}^{\rm out}|V_{\rm back})_i{}_i(V_{\rm back}|V_{\rm scat}^{\rm out})_n\Big|^2 \;,
\end{split}
\ee
where there is a sum over $i=(s'{\bf Q})$ (sum over the spin $s'$ and integral over the momentum ${\bf Q}$), the inner product is given by
\be
(\psi|\varphi)=\int\ud^3{\bf x}\,\psi^\dagger(t,{\bf x})\varphi(t,{\bf x}) \;,
\ee
and all fields are evaluated at some time $t_{\rm in}$ chosen such that the field is negligible for $t<t_{\rm in}$.

\subsection{Numerical approach}\label{Numerical approach}

We solve~\eqref{scatDirac} using a pseudo-spectral approach~\cite{Jiang:2012mfn,Jiang:2013wha,Lv:2018wpn,Jiang:2023hbo,Kohlfurst:2015zxi,Kohlfurst:2017hbd,Kohlfurst:2017git,Kohlfurst:2019mag,Antoine:2019fwz}. The spatial directions are discretized, so that at each moment in time, $\psi(t)$ is an array of size $(g,n_x,n_y,n_z)$, where $g=2$ or $4$ is the number of spinor components and $n_i$ is the number of grid points in the $x_i$ direction. The spatial derivatives are computed by Fourier transforming,
\be
\partial_j\psi=\text{inverse FFT}(-ik_j\text{ FFT}[\psi]) \;.
\ee
We used this approach in~\cite{Torgrimsson:2025pao} with Mathematica on a CPU, finding it to be relatively fast for $(1+1)$D fields and slow but feasible for $(2+1)$D. 

Here we will instead implement this on a GPU. We have found JAX~\cite{jax} to be an incredibly powerful tool for this purpose. It is a computation library that offers high performance with an easy-to-use NumPy-like syntax. Indeed, replacing 
\begin{lstlisting}
import numpy as np
\end{lstlisting}
with
\begin{lstlisting}
import jax.numpy as np
jax.config.update("jax_enable_x64", True)
\end{lstlisting}
will handle many of the necessary changes when transitioning from NumPy to JAX. One difference, though, is that JAX arrays are immutable. There is actually very little JAX-specific syntax needed. One case is the use of \texttt{jax.jit} to make GPU-efficient compilations of functions that are called many times. For example, for the differential solver, we define a function that gives $\partial_t\psi=\text{dPsi}(t,\psi)$ as
\begin{lstlisting}[language=Python]
@jax.jit
def dPsi(t,Psi,args):
    # only NumPy-like syntax below
    ...
\end{lstlisting}

With the spatial discretization, we have an ODE. We could in principle solve it with a fixed time step (e.g. using some Runge-Kutta method), but we have instead found it useful to use the \texttt{diffeqsolve} solver from Diffrax~\cite{diffrax}. The syntax for this is
\begin{lstlisting}
from diffrax import diffeqsolve, ODETerm, Dopri5, PIDController
solver = Dopri5()
controller = PIDController(rtol=1e-5, atol=1e-10)
...
term=ODETerm(dPsi)
solution = diffeqsolve(term,solver,t0=tout,t1=tin,dt0=-1e-1,y0=psiOut,stepsize_controller=controller)
\end{lstlisting}
where \texttt{psiOut} is an array of zeros for $\psi_{\rm scat}(t_{\rm out})=0$ discretized. 

\begin{figure}
    \centering
    \includegraphics[width=.49\linewidth]{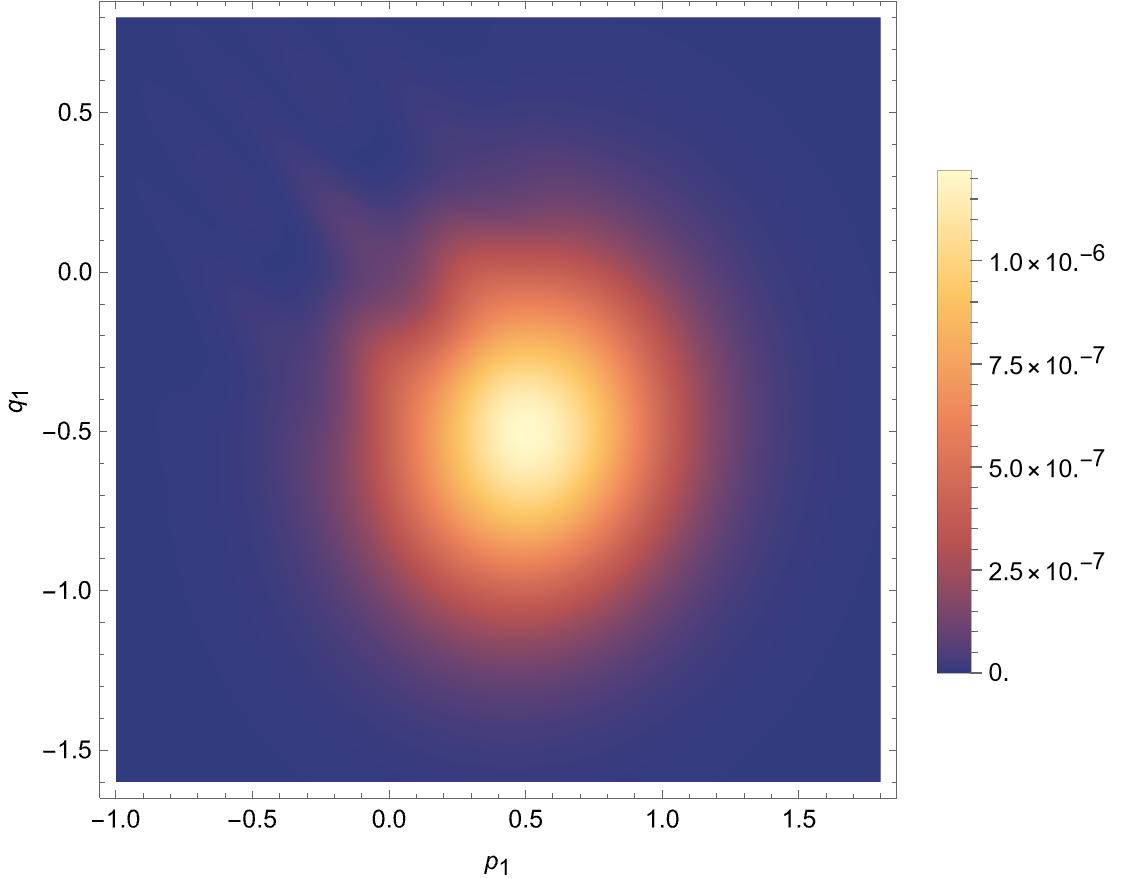}
    \includegraphics[width=.49\linewidth]{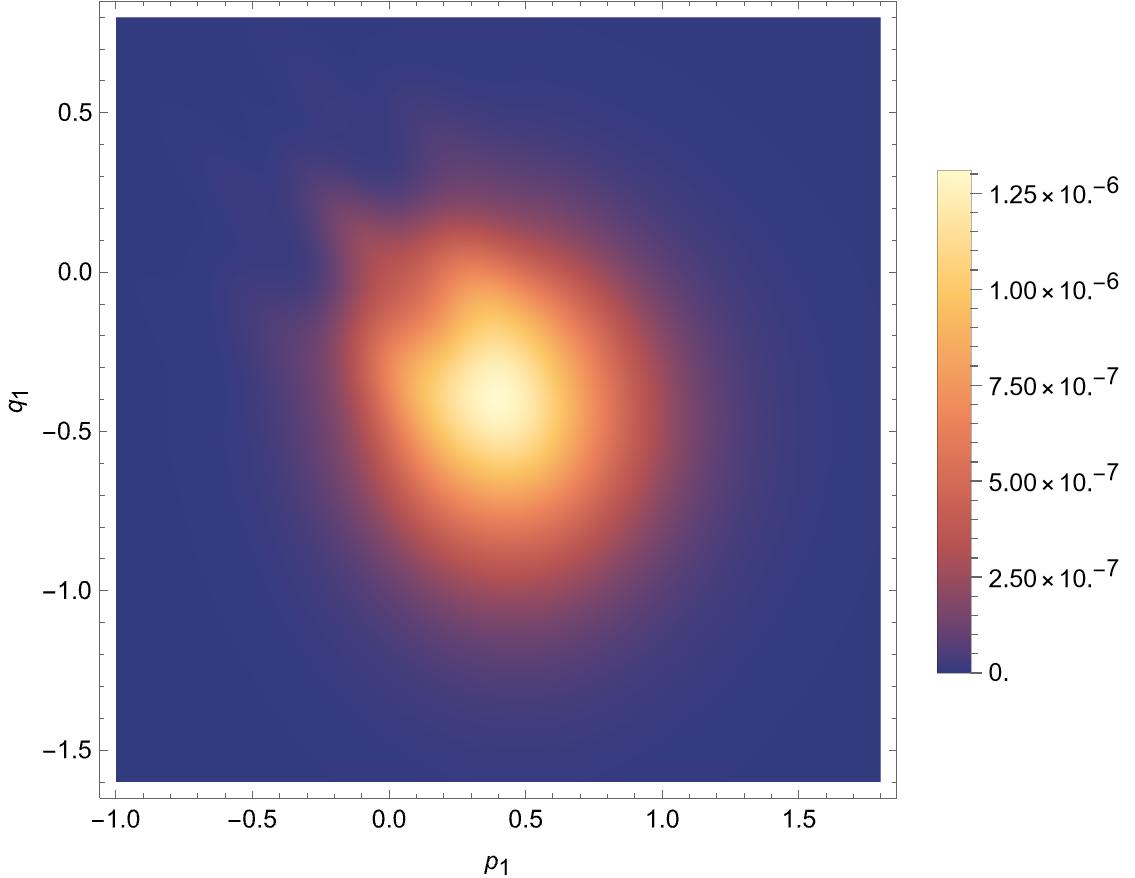}
    \caption{$(2+1)$D. Momentum spectrum with $p_2=p_3=q_2=q_3=0$ for~\eqref{singleE} with $E_0=1/4$, $\omega=E_0$ and $\kappa_x=\kappa_y=E_0/2$. The first row shows the (quadratic) instanton approximation and the second the SWF result. There are three relevant instantons: The one created around $x=0$ gives the dominant contribution, and the two created around $x\approx\pm2.3$ give the interference patterns in the upper-left corner.}
    \label{fig:2plus1single}
\end{figure}

The full code can be found at~\cite{GTGitHub}. For $1+1$ and $2+1$ fields, $N({\bf p},{\bf q})$ only takes seconds to compute for a single value of ${\bf p}$ and ${\bf q}$. One could then compute $N({\bf p},{\bf q})$ on a grid of $({\bf p},{\bf q})$ values, running through the grid points sequentially. For reference, we call this a parallel-sequential approach, since it makes use of the massive parallelization on a GPU for each grid point, but not for running through the grid points.
However, to maximize GPU utilization, we can parallelize the computation on these grid points. For example, for a section of the momentum space where $(p_2, p_3, q_2, q_3)$ are constant and $p_1=-q_1=p$, we can accomplish this using \texttt{jax.vmap} as
\begin{lstlisting}
batched_N = jax.jit(jax.vmap(lambda p: N(p,p2,p3,-p,q2,q3)))
N_values = batched_N(pList)
\end{lstlisting}
where \texttt{N(p1,p2,p3,q1,q2,q3)} solves the Dirac equation and computes the integrals in~\eqref{Nmn} for one point in $({\bf p},{\bf q})$, and \texttt{pList} is an array of \texttt{p} values. For a 2D cross section we could do
\begin{lstlisting}
def inner(p1):
    return jax.vmap(lambda pp1: N(p1, p2, p3, pp1, pp2, pp3))(pList)
    
batched_N = jax.jit(jax.vmap(inner))
N_values = batched_N(pList)
\end{lstlisting}
This parallel-parallel approach is incredibly fast. The time it takes to compute 2D cross sections for the $2+1$ case, like the one in Fig.~\ref{fig:2plus1single}, is measured in seconds, for a grid size of $100\times100$ in $(p_1,q_1)$ and $128\times128$ in $(x,y)$ on a laptop with an NVIDIA GeForce RTX 5070 GPU. Granted, this was for a quite simple field. A field with several oscillations will of course lead to longer runtimes. 

The $3+1$ case, though, is significantly slower, because the extra dimension leads to arrays which take up much more memory. The GeForce 5070 GPU has 8GB in RAM. The A100 GPU available on Google Colab has 80GB, but even for this powerful GPU the code will run out of memory if one tries to apply the above parallel-parallel approach for a $100\times100$ grid. So, for the $3+1$ case in Fig.~\ref{fig:great4Dexample2D} we were forced to make smaller 1D batches, e.g. by running through the $P$ values sequentially and only making batches for the $\Delta$ values. This meant that the $3+1$ case took $\mathcal{O}(1)$ hour to finish for a grid of size $20\times20$. For a field with a couple of oscillations, this would increase further. Thus, going from $2+1$ to $3+1$ takes one to a fundamentally different level of computational challenge. This underscores the need for an approximation method capable of efficiently exploring both momentum and field parameter spaces -- a role naturally suited to the worldline instanton formalism.

\begin{figure}
    \centering
    \includegraphics[width=.49\linewidth]{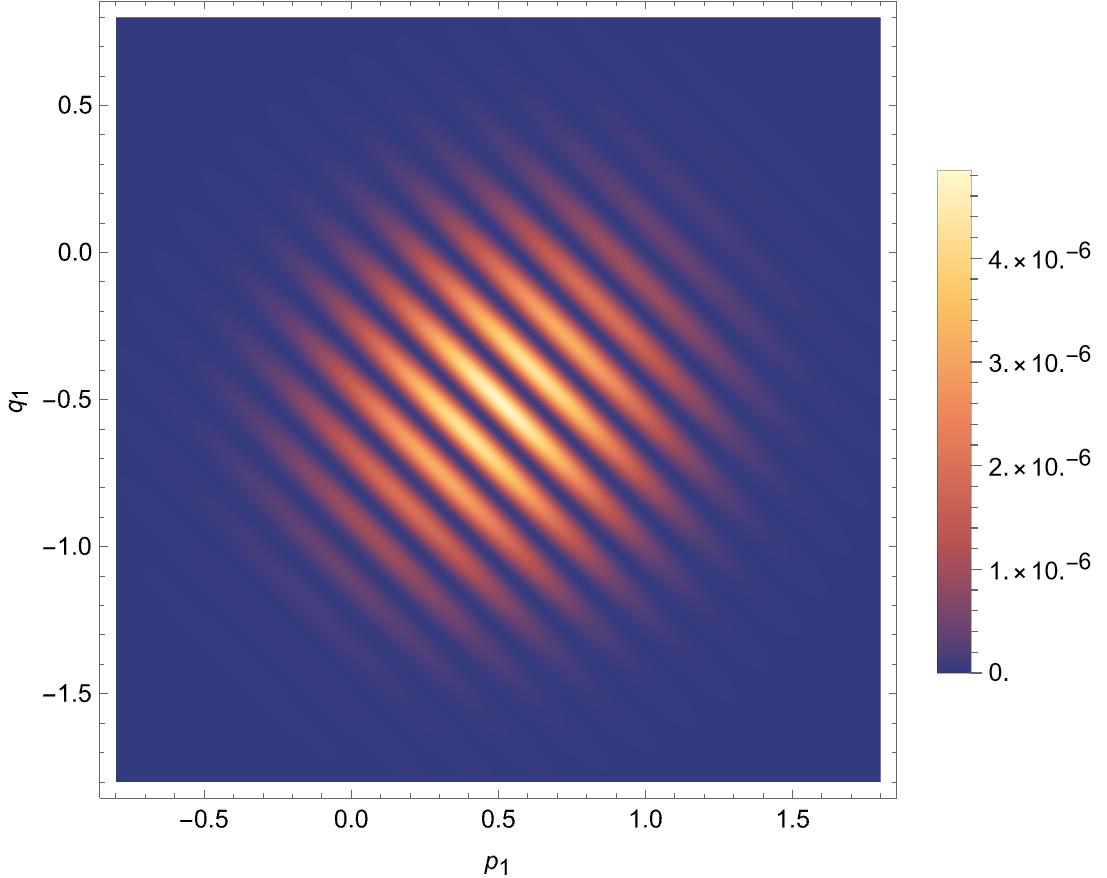}
    \includegraphics[width=.49\linewidth]{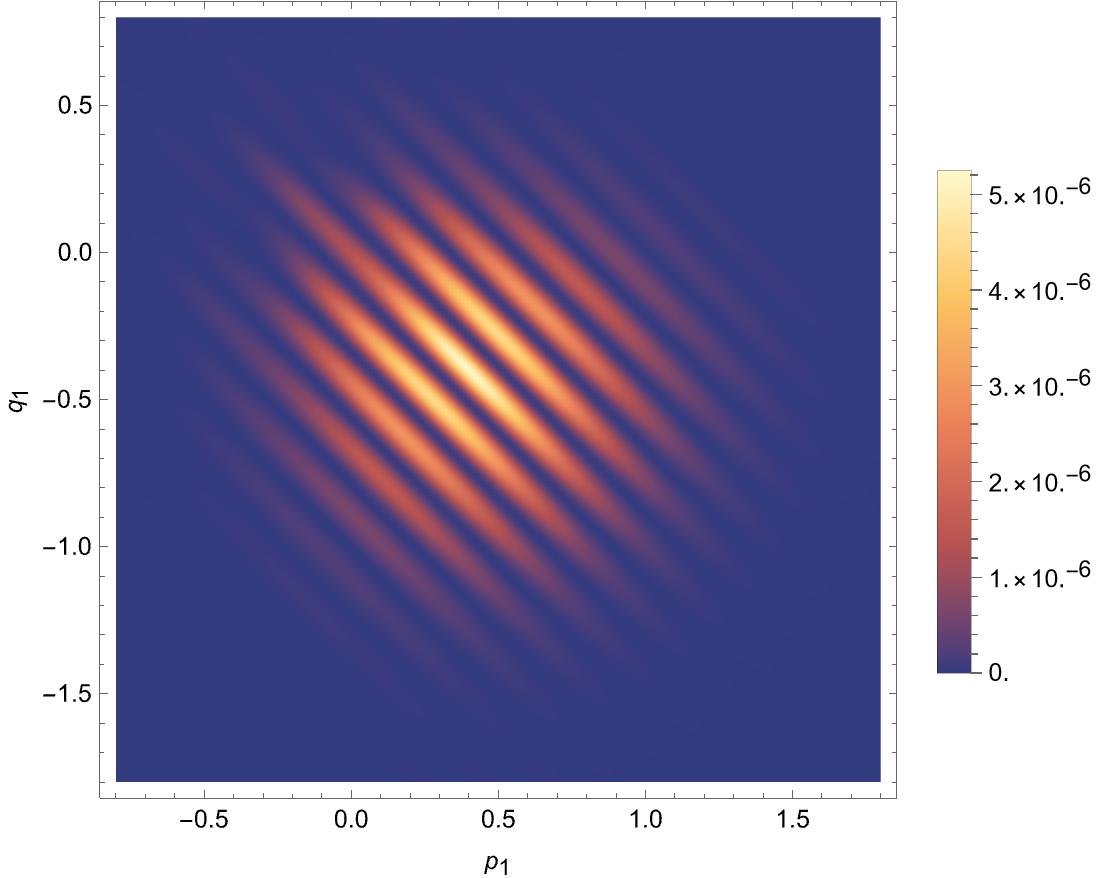}
    \caption{$(2+1)$D. Momentum spectrum with $p_2=p_3=q_2=q_3=0$ for~\eqref{doubleE} with $E_0=1/4$, $\omega=E_0$, $\kappa_x=\kappa_y=E_0/2$ and $\Delta x=1.75/\kappa_x$. The first row shows the (quadratic) instanton approximation and the second the SWF result. We have only included the two dominant instantons.}
    \label{fig:2plus1double}
\end{figure}

\begin{figure}
    \centering
    \includegraphics[width=\linewidth]{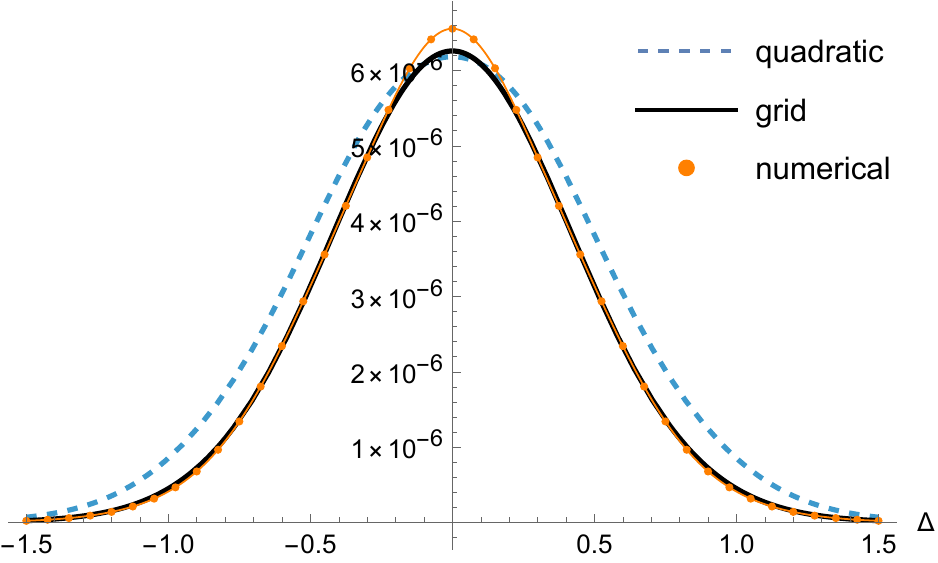}
    \caption{$(3+1)$D. Momentum spectrum with $p_2=p_3=q_2=q_3=0$, $p_1=-P+\frac{\Delta}{2}$, $q_1=P+\frac{\Delta}{2}$, where $P=P_{\rm saddle}\approx0.51$, for~\eqref{singleE} in $3+1$D with $E_0=1/4$, $\omega=E_0$ and $\kappa_x=\kappa_y=\kappa_z=E_0/2$. The numerical SWF points have been computed with a $(x,y,z)$ grid of size $128\times128\times128$. This is an example where the instanton approximations are much better than what one should expect. In general, one should expect relative errors $\mathcal{O}(10\%)$ for $E_0=\mathcal{O}(0.1)$.}
    \label{fig:great4Dexample}
\end{figure}

\begin{figure}
    \centering
    \includegraphics[width=\linewidth]{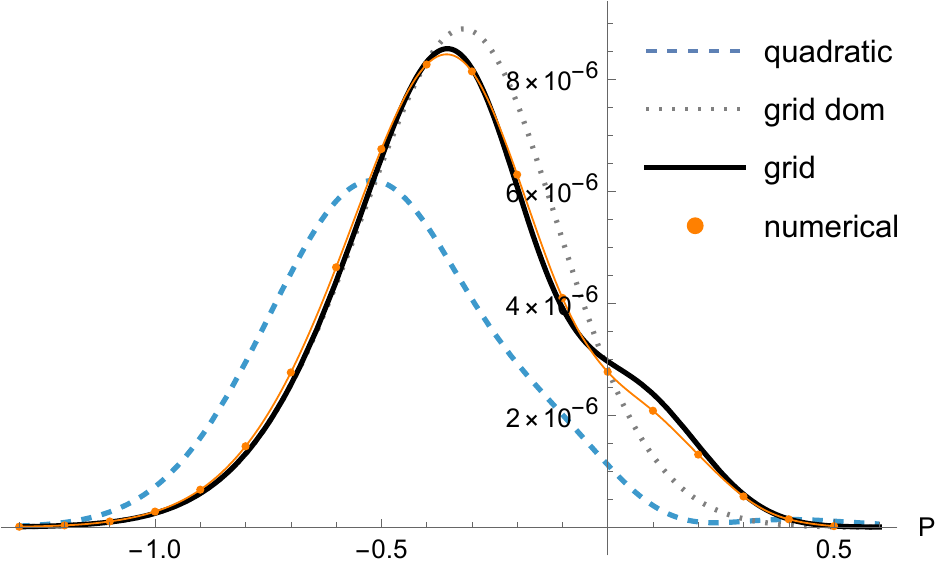}
    \caption{$(3+1)$D. Same as Fig.~\ref{fig:great4Dexample}, but with $\Delta=0$. The ``grid dom'' line shows the grid-instanton approximation including only the dominant instanton, i.e. the one created near $x=0$.}
    \label{fig:great4DexampleP}
\end{figure}

\begin{figure*}
    \centering
    \includegraphics[width=0.24\linewidth]{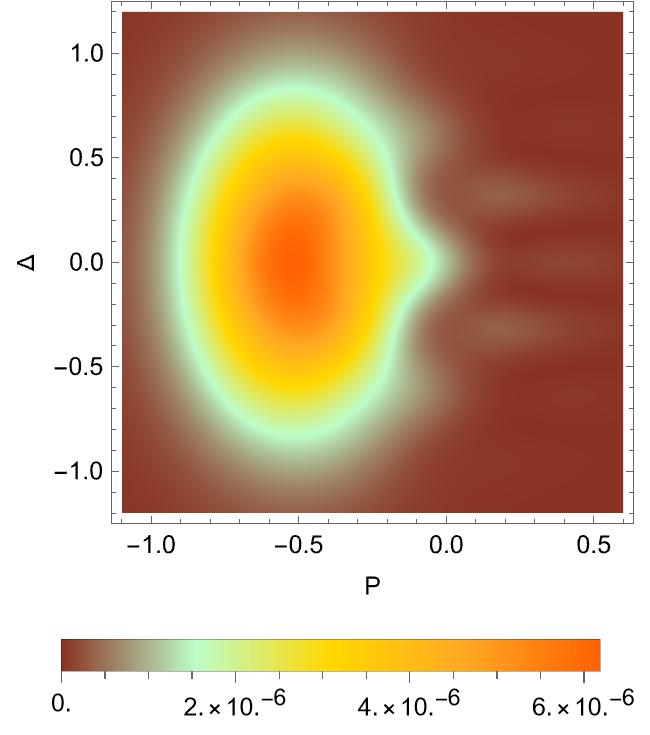}
    \includegraphics[width=0.24\linewidth]{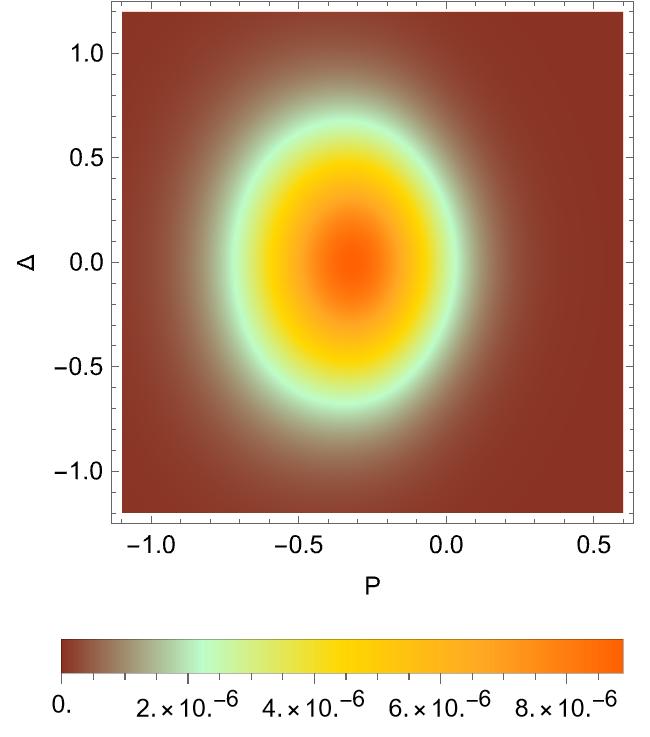}
    \includegraphics[width=0.24\linewidth]{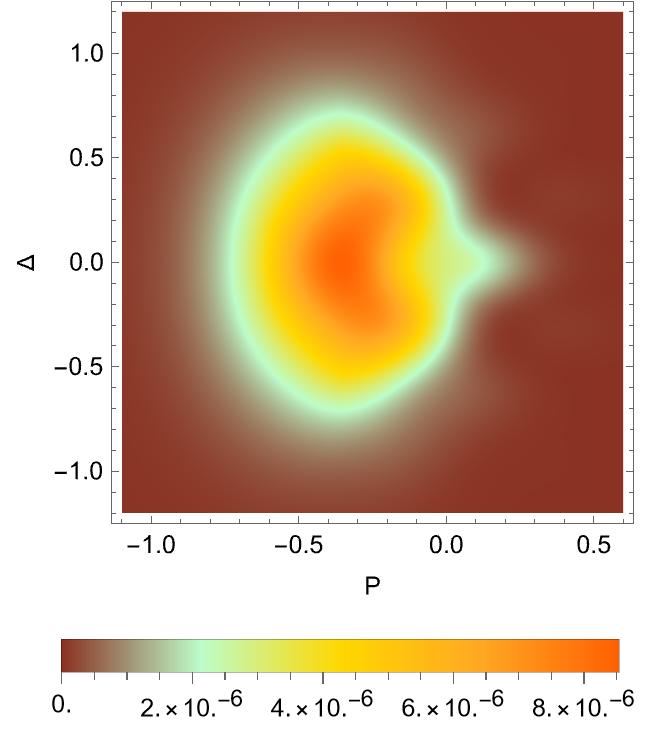}
    \includegraphics[width=0.24\linewidth]{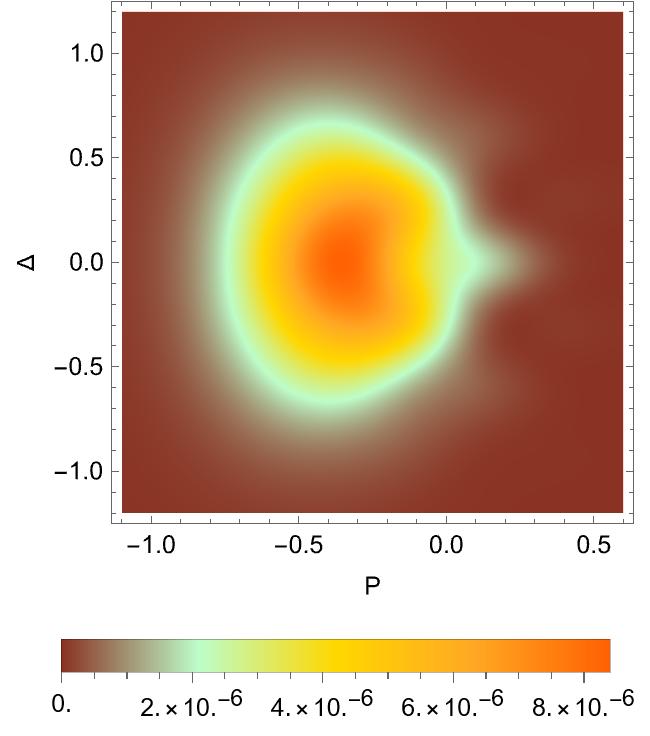}
    \caption{$(3+1)$D. Same as Fig.~\ref{fig:great4Dexample} and~\ref{fig:great4DexampleP}. From left to right: quadratic-instanton approximation, grid-instanton approximation with only the dominant instanton, grid-instanton approximation with the dominant and the two subdominant instantons, and the SWF result.}
    \label{fig:great4Dexample2D}
\end{figure*}

\begin{figure*}
    \centering
    \includegraphics[width=\linewidth]{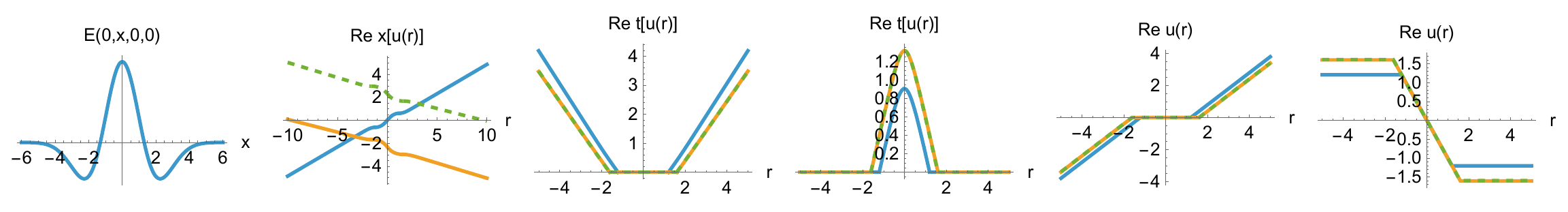}
    \caption{The electric field in~\eqref{singleE} for $t=y=z=0$. $t(u)$ and $x(u)$ ($y(u)=z(u)=0$) are the instantons for the saddle-point values of the (asymptotic) momenta, ${\bf p}_s$ and ${\bf q}_s$. Proper time $u$ follows a complex path, parametrized by a real variable $r$. There are three instantons. The one with $x(0)=0$ is created around the global maximum of the field and hence gives the dominant contribution. The other two are related to each other by symmetry. Since they are created around the two lower field peaks, they give smaller contributions.}
    \label{fig:instantonPlots}
\end{figure*}

\section{Numerical examples}

We consider the following examples:
\be\label{singleE}
\begin{split}
A_0^{\rm one}=&-\frac{E_0}{\kappa_x}\sin(\kappa_x x)\\
\times&\exp\left[-(\omega t)^2-(\kappa_xx)^2-(\kappa_yy)^2-(\kappa_zz)^2\right]
\end{split}
\ee
and
\be\label{doubleE}
A_0^{\rm two}(t,x,y)=A_0^{\rm one}(t,x+\Delta x,y)+A_0^{\rm one}(t,x-\Delta x,y)
\ee
The results are shown in Figs.~\ref{fig:2plus1single}, \ref{fig:2plus1double}, \ref{fig:great4Dexample}, \ref{fig:great4DexampleP} and~\ref{fig:great4Dexample2D}, where we also compare with the instanton approximations. See the appendix for a presentation of the instanton approach. We have two instanton approximations: one in which we expand around the saddle-point values of the momenta, ${\bf p}_s$ and ${\bf q}_s$, which we refer to as the quadratic-instanton approximation. These instantons are shown in~\ref{fig:instantonPlots}. This approximation is very fast, but not as accurate as the grid-instanton approximation, where we have different instantons for each value of ${\bf p}$ and ${\bf q}$.

\section{Nonlinear Breit-Wheeler pair production}

In this section we will show how to use the SWF approach for nonlinear Breit-Wheeler pair production. The amplitude\footnote{See~\cite{unstableVacuumBook} for a careful derivation of the starting point.} separates into three terms,
\be
\begin{split}
M&=\int\ud^4x\,\bar{U}\slashed{\epsilon}e^{-ikx}V \\
&=M_\text{scat,scat}+M_\text{back,scat}+M_\text{scat,back} \;,
\end{split}
\ee
where
\be\label{MabDef}
M_{a,b}=\int\ud^4x\bar{U}_a\slashed{\epsilon}e^{-ikx}V_b 
\ee
and
\be
M_\text{back,back}\propto\delta^4(p+q-k)=0 \;.
\ee
Since all three terms contain $U_{\rm scat}$ and/or $V_{\rm scat}$, the integrands have finite support in ${\bf x}$ for any fixed value of $t$. However, the region of support is the past light cone of the field and so becomes larger at larger $-t$. In particular, the integrands do not vanish at $t<t_{\rm in}$. Instead, the integrals are finite because the integrands oscillate at $t<t_{\rm in}$. We deal with this in two different ways. 

In the first approach, we further separate the amplitude into asymptotic-past and finite-time terms, 
\be
M^{a,b}_{\rm past}=\int_{-\infty}^{t_{\rm in}}\ud t\int\ud^3{\bf x}\,\bar{U}_a\slashed{\epsilon}e^{-ikx}V_b
\ee
and
\be\label{Mabfinite}
M^{a,b}_\text{finite}=\int_{t_{\rm in}}^{t_{\rm out}}\ud t\int\ud^3{\bf x}\,\bar{U}_a\slashed{\epsilon}e^{-ikx}V_b \;.
\ee
We have
\be
M_{\rm future}=\int_{t_{\rm out}}^{\infty}\ud t\int\ud^3{\bf x}\dots=0
\ee
because $U_{\rm scat}=V_{\rm scat}=0$ for $t>t_{\rm out}$. The integrals for $M^{a,b}_\text{finite}$ can be performed directly, since the integrands have finite support for ${\bf x}$ and the time integral is over a finite interval.

To deal with $M^{a,b}_\text{past}$, we Fourier transform the wave functions as
\be\label{UVfourier}
\begin{split}
U_a(t,{\bf x})&=\int\frac{\ud^3{\bf Q}}{(2\pi)^3}U_a(t,{\bf Q})e^{iQ_jx^j}\\
V_a(t,{\bf x})&=\int\frac{\ud^3{\bf Q}}{(2\pi)^3}V_a(t,{\bf Q})e^{iQ_jx^j} \;.
\end{split}
\ee
For the background waves we have simply
\be
\begin{split}
U_{\rm back}(t,{\bf Q})&=(2\pi)^3\delta^3({\bf Q}+{\bf p})u({\bf p})^{-ip_0 t} \\
V_{\rm back}(t,{\bf Q})&=(2\pi)^3\delta^3({\bf Q}-{\bf q})v({\bf p})^{ip_0 t} \;.
\end{split}
\ee
For $t<t_{\rm in}$, the scattered waves become a sum of plane waves,
\be
\begin{split}
U_{\rm scat}(t,{\bf Q})&=U_{\rm scat}^\LCp({\bf Q})e^{iQ_0t}+U_{\rm scat}^\LCm({\bf Q})e^{-iQ_0t} \\
V_{\rm scat}(t,{\bf Q})&=V_{\rm scat}^\LCp({\bf Q})e^{iQ_0t}+V_{\rm scat}^\LCm({\bf Q})e^{-iQ_0t} \;.
\end{split}
\ee
The two terms can be obtained by projecting the full solution as
\be
U_{\rm scat}^\LCpm e^{\pm iQ_0t}=\Lambda_\LCpm U_{\rm scat} \;,
\ee
where $\Lambda_\LCpm$ are two projection matrices
\be\label{LambdaDef}
\Lambda_\LCpm({\bf Q})=\frac{1}{2}\left(1\mp\frac{Q_k\alpha^k+\beta}{Q_0}\right) \;,
\ee
where $\beta=\gamma^0$ and $\alpha^k=\gamma^0\gamma^k$.

Thus, while the integrands do not vanish at $t<t_{\rm in}$, after Fourier transforming, the $t$ dependence is simple and we can perform the $t$ integral analytically,
\be
\begin{split}
M^{a,b}_{\rm past}=\sum_{s,s'=\pm1}\int&\frac{\ud^3{\bf Q}}{(2\pi)^3}\bar{U}_a^s({\bf Q})\slashed{\epsilon}V_b^{s'}({\bf Q}+{\bf k})\\
&\times\frac{e^{i(-sQ_0+s'Q_0'-k_0)t_{\rm in}}}{i(-sQ_0+s'Q_0'-k_0)} \;,
\end{split}
\ee
where $Q_0'=\sqrt{1+({\bf Q}+{\bf k})^2}$. The $t_{\rm in}$ dependence in $M^{a,b}_{\rm past}$ cancels in the sum $M^{a,b}_{\rm past}+M^{a,b}_\text{finite}$.

Thus,
\be\label{MScatScatPast}
\begin{split}
M_\text{scat,scat}^{\rm past}=\sum_{s,s'=\pm1}\int&\frac{\ud^3{\bf Q}}{(2\pi)^3}\bar{U}_{\rm scat}^s({\bf Q})\slashed{\epsilon}V_{\rm scat}^{s'}({\bf Q}+{\bf k})\\
&\times\frac{e^{i(-sQ_0+s'Q_0'-k_0)t_{\rm in}}}{i(-sQ_0+s'Q_0'-k_0)} \;,
\end{split}
\ee
\be
M_\text{back,scat}^{\rm past}=\sum_{s'=\pm1}\bar{u}({\bf p})\slashed{\epsilon}V_{\rm scat}^{s'}({\bf k}-{\bf p})\frac{e^{i(p_0+s'Q_0'-k_0)t_{\rm in}}}{i(p_0+s'Q_0'-k_0)} \;,
\ee
\be
M_\text{scat,back}^{\rm past}=\sum_{s=\pm1}\bar{U}_{\rm scat}^s({\bf q}-{\bf k})\slashed{\epsilon}v_{\rm scat}({\bf q})\frac{e^{i(-sQ_0+q_0-k_0)t_{\rm in}}}{i(-sQ_0+q_0-k_0)} \;.
\ee

In the second approach we use partial integration in $t$. For all $t$, not just for $t<t_{\rm in}$, we Fourier transform as in~\eqref{UVfourier} and split the wave functions as
\be
\begin{split}
U_{\rm scat}(t,{\bf Q})&=U_{\rm scat}^\LCp(t,{\bf Q})e^{iQ_0t}+U_{\rm scat}^\LCm(t,{\bf Q})e^{-iQ_0t} \\
V_{\rm scat}(t,{\bf Q})&=V_{\rm scat}^\LCp(t,{\bf Q})e^{iQ_0t}+V_{\rm scat}^\LCm(t,{\bf Q})e^{-iQ_0t} \;,
\end{split}
\ee
where
\be
\begin{split}
U_{\rm scat}^\LCpm(t,{\bf Q})e^{\pm iQ_0t}&=\Lambda_\LCpm({\bf Q})U_{\rm scat}(t,{\bf Q}) \\
V_{\rm scat}^\LCpm(t,{\bf Q})e^{\pm iQ_0t}&=\Lambda_\LCpm({\bf Q})V_{\rm scat}(t,{\bf Q})\;.
\end{split}
\ee
By performing partial integration in $t$ we obtain
\be\label{MScatScat2}
\begin{split}
M_\text{scat,scat}=\!\sum_{s,s'=\pm1}&\int_{-\infty}^{\infty}\!\ud t\int\frac{\ud^3{\bf Q}}{(2\pi)^3}\frac{ie^{i(-sQ_0+s'Q_0'-k_0)t}}{-sQ_0+s'Q_0'-k_0}\\
&\times\partial_t\left[\bar{U}_{\rm scat}^s(t,{\bf Q})\slashed{\epsilon}V_{\rm scat}^{s'}(t,{\bf Q}+{\bf k})\right] \;,
\end{split}
\ee
\be
\begin{split}
M_\text{back,scat}=\!\sum_{s'=\pm1}&\int_{-\infty}^{\infty}\!\ud t\,\frac{ie^{i(p_0+s'Q_0'-k_0)t}}{p_0+s'Q_0'-k_0}\\
&\times\bar{u}({\bf p})\slashed{\epsilon}\partial_tV_{\rm scat}^{s'}(t,{\bf k}-{\bf p}) \;,
\end{split}
\ee
\be
\begin{split}
M_\text{scat,back}=\!\sum_{s=\pm1}&\int_{-\infty}^{\infty}\!\ud t\,\frac{ie^{i(-sQ_0+q_0-k_0)t}}{-sQ_0+q_0-k_0}\\
&\times\partial_t\bar{U}_{\rm scat}^s(t,{\bf q}-{\bf k})\slashed{\epsilon}v({\bf q}) \;.
\end{split}
\ee
Thanks to $\partial_t[...]$ the integrands now decay quickly to zero, not just as $t\to+\infty$, but also as $t\to-\infty$.

\subsection{Numerical approach}

We solve the Dirac equation to find $U_{\rm scat}$ and $V_{\rm scat}$ in the same way as in~\eqref{Numerical approach}. For Schwinger pair production, only $U_{\rm scat}(t_{\rm in})$ and $V_{\rm scat}(t_{\rm in})$ enter the formula~\eqref{Nmn} for the probability. For Breit-Wheeler, we have integrals over $t$. Storing $U_{\rm scat}(t,{\bf x})$ and $V_{\rm scat}(t,{\bf x})$ at each time step can easily lead to arrays which take up too much memory. We therefore perform the spatial integrals at each time step while solving the Dirac equation, and only store the integrals. For example, for the term in~\eqref{MScatScat2} we compute the $t$ integrand
\be\label{MScatScatGrand}
\begin{split}
\dot{M}_\text{scat,scat}(t)=&\!\sum_{s,s'=\pm1}\int\frac{\ud^3{\bf Q}}{(2\pi)^3}\frac{ie^{i(-sQ_0+s'Q_0'-k_0)t}}{-sQ_0+s'Q_0'-k_0}\\
\times&\partial_t\left[\bar{U}_{\rm scat}^s(t,{\bf Q})\slashed{\epsilon}V_{\rm scat}^{s'}(t,{\bf Q}+{\bf k})\right] 
\end{split}
\ee
at each time step. The time derivative in~\eqref{MScatScatGrand} can then be obtained using the same $\partial_t\psi$ that is anyway computed at each time step for the integration of the Dirac equation. After the Dirac equation has been completely solved, we are left with 1D arrays, e.g. for $[\dot{M}_\text{scat,scat}(t_0),\dot{M}_\text{scat,scat}(t_1),\dots,\dot{M}_\text{scat,scat}(t_n)]$, which can then be integrated over $t$. (If we use the past-finite-split approach, we also need to store $U_{\rm scat}(t_{\rm in})$ and $V_{\rm scat}(t_{\rm in})$ to compute $M_{\rm past}^{a,b}$.) 

The $t$ integrals can also be combined into the integration of the Dirac equation, by appending the 3 functions 
\be
M_{a,b}(t)=\int^t\ud\tilde{t}\,\dot{M}_{a,b}(\tilde{t})
\ee
to the array containing the discretized $\psi(t,{\bf x})$, so that $\dot{Y}=dY(t,Y)$ where $Y=(\psi,M)$ and $dY=(\dot{\psi},\dot{M})$. As a flattened array, $Y$ then has
\be
\underset{\text{re im}}{2}\left(\underset{U V}{2}\times g\times n_x^{D-1}+\underset{M}{3}\right)
\ee
components, where the first $2$ comes from storing the real and imaginary parts separately; the second 2 comes from solving the uncoupled Dirac equations for $U_{\rm scat}$ and $V_{\rm scat}$ together, so that we can compute $M(t)$ at each time step; $g=4$ is the number of spinor components; $n_x$ is the number of grid points in one spatial dimension; $D-1$ the number of nontrivial spatial dimensions; and $[M_{\rm scat,scat}(t),M_{\rm back,scat}(t),M_{\rm scat,back}(t)]$ give an additional 3 complex numbers. As $6\lll n_x^{D-1}$, we are dealing with arrays that are approximately twice the size of those for Schwinger pair production, where there is no need to solve for $U_{\rm scat}$ and $V_{\rm scat}$ at the same time. But this is still much smaller than if we were to store $U_{\rm scat}(t,{\bf x})$ or $V_{\rm scat}(t,{\bf x})$ at each time step, which would involve arrays with    
\be
\underset{\text{re im}}{2}\times g\times n_x^{D-1}n_t
\ee
components, where $n_t$ is the number of time steps.

The next step in the derivation of formulas would be to include a photon wave packet. Doing so is phenomenologically important since one can expect qualitatively different results for a photon wave function that is localized on the scale of the field compared to plane-wave photons. However, we can still use the above formulas for wave packets. Indeed, for $M_{\rm finite}^{a,b}$ we could just replace $\epsilon_\mu e^{-ikx}$ in~\eqref{Mabfinite} with some wave packet $f_\mu(x)$, which will (assuming some nice Gaussian wave packet) just reduce some of the oscillations of the integrands and hence make the ${\bf x}$ integrals easier to compute. For $M_{\rm past}^{\rm scat,scat}$, we can compute~\eqref{MScatScatPast} on a grid in ${\bf k}$ and then integrate the result weighted by the wave packet in momentum space, $f({\bf k})$. Our preliminary tests suggest that one needs much fewer points for this grid compared to the grids we use to solve the Dirac equation. Note also that the photon does not enter at all in the computation of the fermion wave functions, $U$ and $V$. It only enters when we integrate those solutions as in~\eqref{MabDef}. Thus, in this first paper on SWF for BW, we content ourselves with a plane-wave photon. 

\begin{figure}
    \centering
    \includegraphics[width=\linewidth]{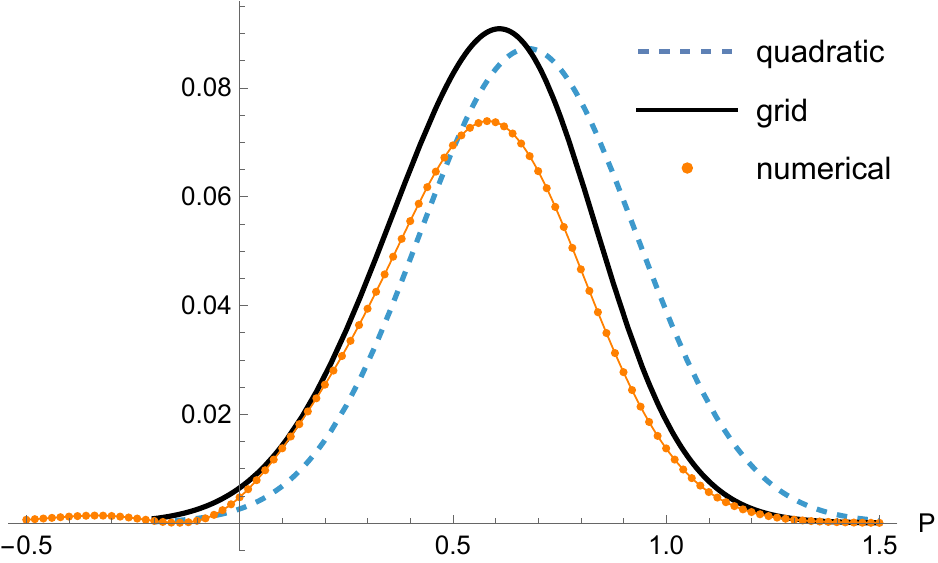}
    \caption{$(1+1)$D. Spectrum, $M/[(2\pi)^2\delta_{y,z}(q+p-k)]$, for nonlinear Breit-Wheeler pair production for the field in~\eqref{singleE} with $-E_0=\omega=1/6$ and $\kappa_x=\omega/2$ ($\kappa_y=\kappa_z=0$), and with momenta $k_1=k_3=q_3=p_3=0$, $k_2=1.2=2q_2=2p_2$ and $p_1=-q_1=P$, spin up in the $x$-basis ($s=+1$ in~\eqref{uv}) and parallel photon polarization ($\epsilon_{(\LCpara)}$ in~\eqref{epsParaPerp}).}
    \label{fig:BWplot}
\end{figure}

While the above formulas are valid for general fields, here we will just consider a $(1+1)$D example. Fig.~\ref{fig:BWplot} shows a 1D cross section of the momentum spectrum and a comparison with the quadratic- and grid-instanton approximations, for which we have only included the instanton ``created'' near the global field maximum, i.e. near $x=0$.  The relative error of these approximations is on the order of magnitude one can expect for $E_0\gtrsim\mathcal{O}(0.1)$. A Jupyter notebook with the code can be found at~\cite{GTGitHub}.

\section{Conclusions}

We have shown how to solve the Dirac equation in $2+1$ and $3+1$ dimensional background fields on a GPU by combining the scattered-wave-function approach~\cite{Torgrimsson:2025pao} with GPU tools such as JAX. We have tested the methods for some simple electric fields with only a few oscillations. Fields with more oscillations will of course lead to larger arrays and longer runtimes, as will smaller values of $\gamma=\omega/E$. We have also seen that, while $(3+1)$D is doable, it is significantly slower than $(2+1)$D, as one can easily run out of memory on even more powerful GPUs. However, we have compared the SWF results with the worldline-instanton approximations and found good agreement, so one can use the latter, which is much faster, to scan the parameter space before improving on the precision of the approximation with a GPU computation of the SWF result. 
With these two methods, one can now start to explore strong-field processes in fully 4D solutions to Maxwell's equation. 

We have also extended the SWF approach to nonlinear Breit-Wheeler pair production. We expect that the extension to nonlinear Compton scattering will be similar. In this paper we considered for simplicity BW by a plane-wave photon. It would be interesting to study the role of photon or fermion wave packets.    

As mentioned in a very recent paper~\cite{JAXinCell}, one can use JAX for automatic differentiation. It would be interesting to try to use that to find the values of ${\bf p}$ and ${\bf q}$ that maximize the pair production probability.

\appendix

\section{Worldline instanton approximation}

In this appendix we will summarize the necessary formulas for the worldline-instanton method in~\cite{DegliEsposti:2021its,DegliEsposti:2022yqw,DegliEsposti:2023qqu,DegliEsposti:2023fbv} and generalize some of the results.

The worldline representation of the dressed propagator is given by
\be\label{propagatorWorldline}
\begin{split}
&S(x_\LCp,x_\LCm)=(i\slashed{\partial}_{x_\LCp}-\slashed{A}(x_\LCp)+1)\int_0^\infty\frac{\ud T}{2}\int\limits_{q(0)=x_\LCm}^{q(1)=x_\LCp}\mathcal{D}x\,\mathcal{P}\\
&\times\exp\left\{-i\left[\frac{T}{2}+\int_0^1\!\ud\tau\left(\frac{\dot{x}^2}{2T}+A\dot{x}+\frac{T}{4}\sigma^{\mu\nu}F_{\mu\nu}\right)\right]\right\} \;,
\end{split}
\ee
where $\sigma^{\mu\nu}=\frac{i}{2}[\gamma^\mu,\gamma^\nu]$ and $\mathcal{P}$ means $\tau$ ordering,
and the probability amplitude is obtained using the Lehmann-Symanzik-Zimmermann (LSZ) reduction formula
\be\label{eq:LSZ3pair}
M=\lim_{t_\LCpm\to\infty}\int \ud^3x_\LCp\ud^3 x_\LCm e^{ipx_\LCp+ip'x_\LCm}
\bar{u}\gamma^0S(x_\LCp,x_\LCm)\gamma^0 v \;.
\ee
The integrals are performed with the saddle-point method.

\subsection{Exponential part}

The instanton, $x^\mu(u)$, is a complex solution to the Lorentz-force equation,
\be\label{LorentzForce}
\ddot{x}^\mu=F^{\mu\nu}(x)\dot{x}_\nu \;,
\ee
with boundary conditions at asymptotic proper times, $u\to\pm u$,
\be\label{asymptoticCondition}
\dot{x}^\mu(-\infty)=-q^\mu \qquad
\dot{x}^\mu(\infty)=p^\mu \;,
\ee
where $q^\mu$ and $p^\mu$ are the (asymptotic) momenta of the positron and electron. Since we have boundary conditions~\eqref{asymptoticCondition} rather than initial conditions, we are forced to use the Newton-Raphson method, where we first make a guess for $x^\mu(0)$ and then solve~\eqref{LorentzForce} a couple of times until we find a $x^\mu(0)$ which gives the correct $\dot{x}^\mu(\pm\infty)$. 

$u$ will generally follow a complex contour. We have the freedom to choose a contour such that $\dot{t}=0$ somewhere in the middle of the instanton, and we can choose this to be the origin of the complex proper-time plane, $\dot{t}(u=0)=0$. It follows from~\eqref{LorentzForce} that $\dot{x}_\mu^2(u)$ is constant, so we can use $\dot{x}_\mu^2(0)=1$ as a condition to further reduce the number of variables to vary in the Newton-Raphson procedure. If we impose the spatial components of~\eqref{asymptoticCondition}, then the temporal components will automatically be fulfilled. The real and imaginary parts of~\eqref{asymptoticCondition} therefore give us $4(D-1)$ real conditions, where $D$ is the number of nontrivial space-time dimensions. This is the same number as the undetermined integration constants: $x^\mu(0)\to 2D$, $\dot{x}^k(0)\to 2(D-1)$, and $\dot{x}_\mu^2(0)=1\to -2$. Thus, the integration constants to vary are $x^\mu(0)$ and
\be
\begin{split}
D&=2: \qquad \dot{x}(0)=\pm i \\
D&=3: \qquad [\dot{x}(0),\dot{y}(0)]=i[\cos\Omega,\sin\Omega] \\
D&=4: \qquad \dot{\bf x}(0)=i[\sin\theta\cos\Omega,\sin\theta\sin\Omega,\cos\theta] \;,
\end{split}
\ee
where $\Omega$ and $\theta$ can be complex. If the field has some symmetry, then the number of undetermined constants can be reduced further. 

There are in general more than one instanton for each $(q^\mu,p^\mu)$. Each instanton gives one term to the amplitude. The exponential part of an amplitude term is given by
\be\label{varphiInt}
\varphi=i\int\ud u\, x^\mu\partial_\mu A_\nu\dot{x}^\nu \;.
\ee

\subsection{Prefactor}

In the computation of the prefactor, we introduce finite proper-time end points, $u_0$ and $u_1$, and $T=u_1-u_0$. $u$ follows a complex contour, starting at $u_0$ and ending at $u_1$.   
Expanding around the instanton, $x_{\rm int. var.}^\mu\to x^\mu+\delta x^\mu$, gives a Gaussian path integral,
\be\label{pathIntRatio}
\int\mathcal{D}\delta x\exp\left\{-\frac{i}{2}\int\ud u\,\delta x\Lambda\delta x\right\}=\frac{1}{(2\pi T)^2}\left[\frac{\text{det}\Lambda_{\rm free}}{\text{det}\Lambda}\right]^{1/2} \;,
\ee
where
\be
\Lambda_{\mu\nu}=-\eta_{\mu\nu}\partial_u^2+F_{\mu\nu}\partial_u+\partial_\nu F_{\mu\rho}\dot{\rho} \;.
\ee
The functional determinant can be computed as an ordinary determinant using the Gelfand-Yaglom method~\cite{Dunne:2006st,Dunne:2006ur},
\be\label{GY}
\text{det }\Lambda=\text{det }\delta x_{(\mu)}^\nu(u_1) \;,
\ee
where $\delta x_{(\mu)}^\nu$ are solutions to the Jacobi equation
\be\label{JacobiEq}
\Lambda_{\mu\nu}\delta x^\nu=0 
\ee
with initial conditions
\be\label{GYinit}
\delta x_{(\mu)}^\nu(u_0)=0
\qquad
\delta\dot{x}_{(\mu)}^\nu(u_0)=\frac{\delta_\mu^\nu}{T} \;.
\ee
With this normalization we have $\text{det }\Lambda_{\rm free}=1$.

For the imaginary part of the effective action, one has closed instantons, and then~\eqref{GY} is convenient as it is. But to obtain the momentum spectrum, we are dealing with open instantons, and we are supposed to take the asymptotic limit, where $\text{Re }(-u_0,u_1,T,t(u_0),t(u_1))\to\infty$. We will therefore analytically extract these divergences from~\eqref{GY} before starting with a numerical computation. The basic idea is the same as in~\cite{DegliEsposti:2022yqw,DegliEsposti:2023qqu}. We choose $\tilde{u}_0$ and $\tilde{u}_1$ to be two points on the $u$ contour that are just large enough that the instanton is outside the field (i.e. $\ddot{x}^\mu\approx0$) before $\tilde{u}_0$ and after $\tilde{u}_1$. The integration from~\eqref{GYinit} at $u_0$ to $\tilde{u}_0$ is trivial, but allows us to replace~\eqref{GYinit} with
\be\label{deltaxDN}
\delta x_{(\mu)}^\nu(u\sim\tilde{u}_0)=\frac{t_0}{Tq_0} D_{(\mu)}^\nu+\frac{1}{T} N_{(\mu)}^\nu \;,
\ee
where we used $\tilde{u}_0-u_0\approx t_0/q_0=t(u_0)/q_0$ and defined a new set of complete solutions,
\be\label{DNinit}
\begin{aligned}
D_{(\mu)}^\nu(\tilde{u}_0)&=\delta_\mu^\nu
&\qquad
\dot{D}_{(\mu)}^\nu(\tilde{u}_0)&=0 \\
N_{(\mu)}^\nu(\tilde{u}_0)&=0
&\qquad
\dot{N}_{(\mu)}^\nu(\tilde{u}_0)&=\delta_\mu^\nu \;.
\end{aligned}
\ee

Most of the $\delta x$'s grow as $\delta x(u)\sim\delta\dot{x}(\tilde{u}_1)(u-\tilde{u}_1)$ after $\tilde{u}_1$. However, the instanton velocity is also a solution to~\eqref{JacobiEq}, and $\delta x^\mu=\dot{x}^\mu$ gives $\delta\dot{x}^\mu=0$ at both $\tilde{u}_0$ and $\tilde{u}_1$, so $\dot{x}^\mu$ is a superposition of the $D_{(\nu)}^\mu$ solutions which converges to a constant at/after $\tilde{u}_1$. This explains why we need to include the $N_{(\mu)}^\nu$ term in~\eqref{deltaxDN}, even though its prefactor is $\mathcal{O}(1/T)$ compared to the $D_{(\mu)}^\nu$ term.   

We therefore have two contributions to the determinant to leading order in $1/T$. The first one is given by
\be
\begin{split}
D_1&=\left(\frac{t_0}{Tq_0}\right)^D\text{det }(D_0, D_1,\dots,D_{D-1})(u_1)\\
&=:\left(\frac{t_0}{Tq_0}\right)^D\left(\frac{t_1}{p_0}\right)^{D-1}h(\tilde{u}_1) \;,
\end{split}
\ee
where we used $u_1-\tilde{u}_1\approx t_1/p_0=t(u_1)/p_0$ and $D$ is the number of nontrivial space-time coordinates. We have factored out $[t_1/p_0]^{D-1}$ rather than $[t_1/p_0]^D$ because one linear combination of the $D_{(\mu)}$'s is given by $\dot{x}^\mu$, which makes $h$ finite in the asymptotic limit. To make this explicit, we can replace one of the $D_{(\mu)}$'s with $\dot{x}^\mu$, e.g.
\be
D_{(0)}^\mu(u)=\frac{1}{\dot{t}(\tilde{u}_1)}\dot{x}^\mu+\sum_{j=1}^{D-1} a_{(j)}D_{(j)}^\mu(u) \;,
\ee
and then
\be\label{hNum}
h=\frac{1}{-q_0}\text{det}\left[p^{\mu_0},\dot{D}_{(1)}^{\mu_1},\dots,\dot{D}_{(D-1)}^{\mu_{D-1}}\right] \;.
\ee

The second contribution is given by
\be
\begin{split}
D_2&=\left(\frac{t_0}{Tq_0}\right)^{D-1}\frac{1}{T}\bigg[\text{det }(N_0, D_1,\dots,D_{D-1})(u_1)\\
&\hspace{2cm}+\text{det }(D_0, N_1,\dots,D_{D-1})(u_1)+\dots\\
&\hspace{2cm}+\text{det }(D_0, D_1,\dots,N_{D-1})(u_1)\bigg]\\
&=:\left(\frac{t_0}{q_0}\right)^{D-1}\left(\frac{t_1}{p_0}\right)^Dg(\tilde{u}_1) \;.
\end{split}
\ee
In the asymptotic limit, $g$ is finite and given by 
\be\label{gNum}
\begin{split}
g&=\text{det }(\dot{N}_0,\dot{D}_1,\dots,\dot{D}_{D-1})(\tilde{u}_1)\\
&+\text{det }(\dot{D}_0,\dot{N}_1,\dots,\dot{D}_{D-1})(\tilde{u}_1)+\dots
\end{split}
\ee
In~\cite{DegliEsposti:2022yqw}, we showed that $h=g$ for $D=2$. And in~\cite{DegliEsposti:2023qqu} we showed it for a class of symmetric fields for $D=4$. In any case, it is straightforward to check whether $\eqref{hNum}=\eqref{gNum}$ numerically for any field. Assuming this, we have
\be\label{detLambdaSchwinger}
\text{det }\Lambda=D_1+D_2=\left(\frac{t_0t_1}{Tq_0p_0}\right)^{D-1}h(\tilde{u}_1) \;,
\ee
where we used
\be
T=u_1-u_0\to\frac{t_0}{q_0}+\frac{t_1}{p_0} \;.
\ee

The integrals over ${\bf X}=\{T,{\bf x}(u_0),{\bf x}(u_1)\}$ can be perform as in~\cite{DegliEsposti:2022yqw,DegliEsposti:2023qqu}. We find
\be
\int\ud^{2D-1}\delta{\bf X}e^{-\delta{\bf X}\cdot{\bf H}\cdot\delta{\bf X}}=\frac{\pi^{(2D-1)/2}}{\sqrt{\text{det }{\bf H}}} \;,
\ee
where, up to a phase that is independent of the instanton,
\be
\text{det }{\bf H}=\frac{(q_0p_0)^{D+1}}{2^{2D-1}(t_0t_1)^{D-1}T}
\ee

As explained in~\cite{DegliEsposti:2022yqw}, each trivial dimension ($l$) gives a factor of
\be
\sqrt{2\pi T}2\pi\delta(p'_l+p_l) \;.
\ee

\subsection{Final results}

Collecting everything gives\footnote{This assumes a single instanton. Otherwise we have to sum the instanton terms on the amplitude level as in~\cite{DegliEsposti:2024upq}.}
\be\label{P11}
D=1+1:\qquad P=V_yV_z\int\frac{\ud^3p\ud q_1}{(2\pi)^3p_0q_0}\frac{2\mathcal{S}}{|h|}e^{-\mathcal{A}} \;,
\ee
\be\label{P21}
D=2+1:\qquad P=V_z\int\frac{\ud^3p\ud q_1\ud q_2}{(2\pi)^3p_0q_0}\frac{2\mathcal{S}}{|h|}e^{-\mathcal{A}} \;,
\ee
\be\label{P31}
D=3+1:\qquad P=\int\frac{\ud^3p\ud^3q}{(2\pi)^3p_0q_0}\frac{2\mathcal{S}}{|h|}e^{-\mathcal{A}} \;,
\ee
where the exponent is given by the real part of~\eqref{varphiInt},
\be\label{expA}
\mathcal{A}=2\text{Im}\int\ud u\, x^\mu\partial_\mu A_\nu\dot{x}^\nu \;,
\ee
$V_l$ is a volume factor for the trivial dimension $l$, $h$ is given by~\eqref{hNum}, and $\mathcal{S}$ is a spin term. For the type of fields considered in~\cite{DegliEsposti:2022yqw,DegliEsposti:2023qqu}, we have $\mathcal{S}=1$ after summing over the spins. Eq.~\eqref{P11} was obtained in~\cite{DegliEsposti:2022yqw}, and~\eqref{P11} was obtained in~\cite{DegliEsposti:2023qqu} for the special case where the field is 4D but the instanton only sees a 2D field\footnote{Note that, compared to the notation in~\cite{DegliEsposti:2023qqu}, we have $h_{\rm here}=h_{\rm there}(\bar{\phi}')^2$.}.   

The above expressions give the leading order in a weak-field expansion, and should therefore have a simple dependence on $E_0$, where $F_{\mu\nu}(x^\rho)=E_0\hat{F}_{\mu\nu}(E_0\gamma x^\rho)$ and $\gamma=\omega/E_0=\mathcal{O}(E_0^0)$. We can see this by rescaling $x^\mu\to x^\mu/E_0$ and $u\to u/E_0$, which removes $E_0$ from the equations of motion~\eqref{LorentzForce} and~\eqref{JacobiEq} and shows that $\mathcal{A}(E_0,\gamma)=\mathcal{A}(\gamma)/E_0$ and $h(E_0,\gamma)=E^{D-1}h(\gamma)$.

\subsection{Expansion around momentum saddle points}

We can solve the Lorentz-force equation for any values of ${\bf q}$ and ${\bf p}$, plug the solution(s) into~\eqref{expA} and obtain the probability spectrum on a grid of ${\bf q}$, ${\bf p}$ values. But it is much faster to expand around the saddle point values, ${\bf q}_s$ and ${\bf p}_s$. We refer to the results as the grid and quadratic instanton approximations.
As explained in~\cite{DegliEsposti:2022yqw,DegliEsposti:2023qqu}, the first derivatives of the exponent are given by simple formulas that only involve the instanton (i.e. no $\delta x$),
\be\label{firstOrderDer}
\begin{split}
\frac{\partial\varphi}{\partial q^k}&=-i\left(x^k-\frac{q^k}{q_0}t\right)(\tilde{u}_0)\\
\frac{\partial\varphi}{\partial p^k}&=-i\left(x^k-\frac{p^k}{p_0}t\right)(\tilde{u}_1) \;.
\end{split}
\ee
For the second derivatives we need $\partial x^\mu/\partial q^k$ and $\partial x^\mu/\partial p^k$, which we obtain as\footnote{The sign is just to use the same convention as in~\cite{DegliEsposti:2022yqw,DegliEsposti:2023qqu}, where we used $q_k$ rather than $q^k=-q_k$.}
\be
\begin{split}
\frac{\partial x^\mu}{\partial q^k}&=\frac{x^\mu(q^k+\ud q^k)-x^\mu(q^k)}{\ud q^k}=-\delta x_{[k]}^\mu \\
\frac{\partial x^\mu}{\partial p^k}&=\frac{x^\mu(p^k+\ud p^k)-x^\mu(p^k)}{\ud p^k}=-\delta x_{\{k\}}^\mu\;,
\end{split}
\ee
where $\delta x_{[k]}^\mu$ and $\delta x_{\{k\}}^\mu$ are solutions to the first-order perturbation around the Lorentz-force equation,
\be\label{firstOrderLorentz}
\delta\ddot{x}^\mu=F^{\mu\nu}\delta\dot{x}_\nu+\partial_\nu F^{\mu\rho}\dot{x}_\rho\delta x^\nu \;,
\ee
with boundary conditions obtained by differentiating~\eqref{asymptoticCondition},
\be\label{deltaBC}
\begin{aligned}
\delta\dot{x}_{[l]}^k(-\infty)&=\delta_{kl}
&\qquad
\delta\dot{x}_{[l]}^k(\infty)&=0 \\
\delta\dot{x}_{\{l\}}^k(-\infty)&=0
&\qquad
\delta\dot{x}_{\{l\}}^k(\infty)&=-\delta_{kl} \;.
\end{aligned}
\ee
Differentiating~\eqref{firstOrderDer} gives
\be
\begin{split}
\frac{\partial^2\varphi}{\partial q^k\partial q^l}&=i\left(\delta x_{[l]}^k-\frac{q^k}{q_0}\delta t_{[l]}+\frac{q_0^2\delta_{kl}-q^kq^l}{q_0^3}t\right)(\tilde{u}_0)\\
&=i\left(\delta x_{[k]}^l-\frac{q^l}{q_0}\delta t_{[k]}+\frac{q_0^2\delta_{kl}-q^kq^l}{q_0^3}t\right)(\tilde{u}_0)
\end{split}
\ee
\be
\begin{split}
\frac{\partial^2\varphi}{\partial p^k\partial p^l}&=i\left(\delta x_{\{l\}}^k-\frac{p^k}{p_0}\delta t_{\{l\}}+\frac{p_0^2\delta_{kl}-p^kp^l}{p_0^3}t\right)(\tilde{u}_1)\\
&=i\left(\delta x_{\{k\}}^l-\frac{p^l}{p_0}\delta t_{\{k\}}+\frac{p_0^2\delta_{kl}-p^kp^l}{p_0^3}t\right)(\tilde{u}_1)
\end{split}
\ee
\be
\begin{split}
\frac{\partial^2\varphi}{\partial q^k\partial p^l}&=i\left(\delta x_{\{l\}}^k-\frac{q^k}{q_0}\delta t_{\{l\}}\right)(\tilde{u}_0)\\
&=i\left(\delta x_{[k]}^l-\frac{p^l}{p_0}\delta t_{[k]}\right)(\tilde{u}_1) \;.
\end{split}
\ee

But~\eqref{firstOrderLorentz} is the same Jacobi equation as in~\eqref{JacobiEq} for the computation of $\text{det }\Lambda$. Thus, (to this order in the weak-field expansion) we only have two equations to solve: the Lorentz-force equation and the Jacobi equation. For the Jacobi equation, we have initial conditions~\eqref{GYinit} for $\delta x_{(\mu)}^\nu$ for $\text{det }\Lambda$, but Neumann conditions~\eqref{deltaBC} at two different proper times for $\delta x_{[k]}^\mu$ and $\delta x_{\{k\}}^\mu$. These Neumann conditions do not mean that we need to use the Newton-Raphson method to solve the Jacobi equation, because it is linear and homogenous, so any solution to~\eqref{firstOrderLorentz} can be expressed as a superposition of the solutions in~\eqref{DNinit}, or in terms of $\mathcal{D}_{(\mu)}^\nu$ and $\mathcal{N}_{(\mu)}^\nu$ with initial conditions at $u=0$,
\be\label{D0N0init}
\begin{aligned}
\mathcal{D}_{(\mu)}^\nu(0)&=\delta_\mu^\nu
&\qquad
\dot{\mathcal{D}}_{(\mu)}^\nu(0)&=0 \\
\mathcal{N}_{(\mu)}^\nu(0)&=0
&\qquad
\dot{\mathcal{N}}_{(\mu)}^\nu(0)&=\delta_\mu^\nu \;.
\end{aligned}
\ee
The $(\mathcal{D},\mathcal{N})$ basis is convenient because we start the numerical integration of the Lorentz-force equation at $u=0$ and integrate out to $\tilde{u}_0$ and $\tilde{u}_1$, which are automatically determined by the numerical solver to be those points where e.g. $|t''(u)|$ has become smaller than some error tolerance. After the basis solutions $(\mathcal{D},\mathcal{N})$ have been obtained, the calculation of the coefficients in
\be
\delta x^\mu(u)=\sum_\nu[\alpha_{(\nu)}\mathcal{D}_{(\nu)}^\mu(u)+\beta_{(\nu)}\mathcal{N}_{(\nu)}^\mu(u)]
\ee
becomes a simple linear-algebra problem.

We often find it useful to use the following linear combination of the electron and positron momenta,
\be
q^k=P^k+\frac{\Delta^k}{2}
\qquad
p^k=-P^k+\frac{\Delta^k}{2} \;.
\ee

\subsection{Spin part}

The spinor part of~\eqref{propagatorWorldline} can be expressed in terms of $\mathcal{E}(\infty,-\infty)$, where
\be\label{Edef}
\mathcal{E}(u,u_0)=(\slashed{\pi}(u)+1)\mathcal{P}
\exp\left\{-\frac{i}{4}\int_{u_0}^u\sigma^{\mu\nu}F_{\mu\nu}\right\} \;,
\ee
where $\pi^\mu(u)=\dot{x}^\mu(u)$. From
\be\label{spinODE}
\partial_u\mathcal{E}(u,u_0)=-\frac{i}{4}\sigma^{\mu\nu}F_{\mu\nu}(x[u])\mathcal{E}(u,u_0) 
\ee
and $\mathcal{E}(u_0,u_0)=\slashed{\pi}(u_0)+1$ we find that $\mathcal{E}$ can also be expressed as 
\be\label{Edef2}
\mathcal{E}(u,u_0)=\mathcal{P}
\exp\left\{-\frac{i}{4}\int_{-\infty}^u\sigma^{\mu\nu}F_{\mu\nu}\right\}(\slashed{\pi}(u_0)+1) \;.
\ee

We showed in~\cite{DegliEsposti:2022yqw,DegliEsposti:2023fbv} that for $E(t,x)$ one can actually perform the $u$ integral in~\eqref{Edef} analytically for an arbitrary pulse, thanks to
\be
E(t[u],x[u])=\pm\frac{\ud}{\ud u}\ln[\pm\dot{t}+\dot{x}] \;,
\ee
which gives
\be\label{sigmaFint}
-\frac{i}{4}\int_{-\infty}^u\sigma^{\mu\nu}F_{\mu\nu}=i\epsilon\sqrt{\frac{q^0-q^1}{p^0-p^1}}\gamma^0\gamma^1 \;,
\ee
where $\epsilon=\pm1$ depending on the sign of $E$. Another way to obtain~\eqref{sigmaFint} is to plug
\be
-\frac{i}{4}\int_{-\infty}^u\sigma^{\mu\nu}F_{\mu\nu}=L\gamma^0\gamma^1
\ee
into~\eqref{Edef} and~\eqref{Edef2} and demand that $\eqref{Edef}=\eqref{Edef2}$. This gives an equation for $L$ whose solution is $L^2=-(q^0-q^1)/(p^0-p^1)$.

We can choose a spin basis as
\be\label{uv}
u_s({\bf p})=\frac{(1+\slashed{p})R_s}{\sqrt{2p_0(p_0+p_1)}}
\quad
v_s({\bf p})=\frac{(1-\slashed{p})R_s}{\sqrt{2p_0(p_0+p_1)}} \;,
\ee
where $s=\pm1$ and
\be\label{Reqs}
\gamma^0\gamma^1R_s=R_s
\qquad
i\gamma^2\gamma^3 R_s=sR_s 
\qquad
R_r^\dagger R_s=\delta_{rs} \;.
\ee

For $E(t,x)$, this gives
\be\label{Msign}
\frac{1}{2\sqrt{q_0p_0}}\bar{u}_r\gamma^0\mathcal{E}\gamma^0 v_s=-i\epsilon\delta_{rs} \;,
\ee
so the spin term in~\eqref{P11}-~\eqref{P31} is given by
\be
\mathcal{S}_{rs}:=\frac{1}{8p_0q_0}|\bar{u}_r\gamma^0\mathcal{E}\gamma^0 v_s|^2=\frac{1}{2}\delta_{rs} 
\ee
or
\be
\mathcal{S}:=\sum_{r,s=\pm1}\mathcal{S}_{rs}=1 \;.
\ee
For arbitrary $F_{\mu\nu}$, $u_r({\bf p})$ and $v_s({\bf q})$, we can compute the spin part by solving~\eqref{spinODE} numerically.

\subsection{Nonlinear Breit-Wheeler pair production}

We can use similar methods for nonlinear Breit-Wheeler pair production, as shown in~\cite{DegliEsposti:2023fbv}.
A common trick~\cite{Strassler:1992zr,McKeon:1994hd,Shaisultanov:1995tm,Dittrich:2000wz,Schubert:2000yt,Schubert:2001he,Ahmad:2016vvw,Gies:2011he} to include a (high-energy) incoherent photon is to replace the coherent background field as
\be
A_\mu(x)\to A_\mu(x)+\epsilon_\mu e^{-ikx} 
\ee
and then select those terms in the amplitude which are linear in the polarization vector $\epsilon_\mu$. 

The instanton is still determined by the asymptotic conditions in~\eqref{asymptoticCondition} and solves the Lorentz-force equation~\eqref{LorentzForce}, except at a single point in the middle of the instanton where the velocity changes discontinuously due to the absorbed high-energy photon,
\be
0<\delta\ll1:\qquad
\dot{x}^\mu(+\delta)-\dot{x}^\mu(-\delta)=k^\mu \;.
\ee
The formula for the exponential part of the amplitude is still given by~\eqref{varphiInt}.

A trivial generalization of the derivation in Appendix C in~\cite{DegliEsposti:2023fbv} gives for the functional determinant
\be\label{detLambdaBW}
\text{det }\Lambda=\left(\frac{t_0t_1}{Tq_0p_0}\right)^Dh(\tilde{u}_1) \;,
\ee
where
\be\label{hNumBW}
h=\text{det}\left[\dot{D}_{0}^{\mu_0},\dots,\dot{D}_{(D-1)}^{\mu_{D-1}}\right] \;,
\ee
and $D_{(\mu)}$ are the Dirichlet solutions~\eqref{DNinit} of the Jacobi equation~\eqref{JacobiEq}.

The integrals over the ordinary integrals, ${\bf X}=(T,\sigma,x_\LCm,...,x_\LCp,...)$, can be performed as described in Appendix D in~\cite{DegliEsposti:2023fbv}. We only include in ${\bf X}$ the components of ${\bf x}_\LCpm$ on which the field depends. With $\delta{\bf X}={\bf X}-{\bf X}_{\rm saddle}$ we find
\be\label{Xint}
\int\ud^{2D}\delta{\bf X}\,e^{-\delta{\bf X}\cdot{\bf H}\cdot\delta{\bf X}}=\frac{\pi^D}{\sqrt{\text{det }{\bf H}}} \;,
\ee
where
\be
\text{det }{\bf H}=\frac{(q_0p_0)^{2+D}T^2}{(4t_\LCp t_\LCm)^D} \;.
\ee

For each trivial dimension $l$ ($\partial_{x^l} F_{\mu\nu}=0$) we change variables as
\be
x_\LCpm^l=\varphi^l\pm\frac{\theta^l}{2} 
\ee
and find
\be\label{varphiIntDelta}
\int\ud^{4-D}{\bm\varphi}e^{i(q+p-k)_l\varphi^l}=(2\pi)^{4-D}\delta^{4-D}(q+p-k) 
\ee
and
\be\label{thetaInt}
\int\ud^{4-D}{\bm\theta}\to(2\pi T)^{(4-D)/2} \;.
\ee

The photon has momentum ${\bf k}$ and polarization vector $\epsilon_\mu=(0,{\bm\epsilon})$. With ${\bf e}$ a unit vector pointing e.g. in the direction of the field, we choose a polarization basis as
\be\label{epsParaPerp}
{\bm\epsilon}_{(\LCperp)}=\frac{{\bf e}\times{\bf k}}{|{\bf e}\times{\bf k}|}
\qquad
{\bm\epsilon}_{(\LCpara)}=\frac{{\bm\epsilon}_{(\LCperp)}\times{\bf k}}{|{\bm\epsilon}_{(\LCperp)}\times{\bf k}|} \;,
\ee
so that $({\bm\epsilon}_{(\LCpara)},{\bm\epsilon}_{(\LCperp)},{\bf k}/|{\bf k}|)$ form a right-handed set of basis vectors. We should replace~\eqref{Edef} and~\eqref{Edef2} with
\be\label{EdefPol}
\begin{split}
\mathcal{E}=&(\slashed{p}+1)\mathcal{P}\bigg[\exp\left\{-\frac{i}{4}\int_0^\infty\!\!\sigma F\right\}\\
\times&\left(\epsilon\dot{x}(0)+\frac{\slashed{k}\slashed{\epsilon}}{2}\right)\exp\left\{-\frac{i}{4}\int_{-\infty}^0\!\!\sigma F\right\} \bigg] 
\end{split}
\ee
and
\be\label{EdefPol2}
\begin{split}
\mathcal{E}=&\mathcal{P}\bigg[\exp\left\{-\frac{i}{4}\int_0^\infty\!\!\sigma F\right\}\\
\times&\left(\epsilon\dot{x}(0)+\frac{\slashed{k}\slashed{\epsilon}}{2}\right)\exp\left\{-\frac{i}{4}\int_{-\infty}^0\!\!\sigma F\right\} \bigg](1-\slashed{q}) \;.
\end{split}
\ee

The two exponentials in~\eqref{EdefPol} can be computed numerically for an arbitrary field using~\eqref{spinODE}. As a simple yet relevant example, we consider $F_{01}=E(t,x)$, $k^1=k^3=q^3=p^3=0$ and
\be
k^2=k^0=2q^2=2p^2=:2\rho
\ee
Writing
\be
-\frac{i}{4}\int_{-\infty}^0\ud u\, E=L_0
\quad
-\frac{i}{4}\int_0^\infty\ud u\, E=L_1
\ee
and demanding that $\eqref{EdefPol}=\eqref{EdefPol2}$ for $\epsilon_{(\LCperp)}$ and $\epsilon_{(\LCpara)}$ allows us to determine
\be
\dot{x}(0)=i\epsilon
\quad
L_0=\sqrt{\frac{q^2-i\epsilon}{q^0+q^1}}
\quad
L_1=\sqrt{\frac{p^2-i\epsilon}{p^0-p^1}} \;,
\ee
where $\epsilon=\pm1$. Plugging these solutions into~\eqref{EdefPol} or~\eqref{EdefPol2} gives for $\epsilon_{(\LCperp)}$
\be\label{Mperp}
\frac{1}{2\sqrt{q_0p_0}}\bar{u}_r\gamma^0\mathcal{E}\gamma^0 v_s=-\frac{\rho}{\sqrt{1+\rho^2}}
\begin{pmatrix}\epsilon&1\\-1&-\epsilon\end{pmatrix}
\ee
and for $\epsilon_{(\LCpara)}$
\be\label{Mpara}
\frac{1}{2\sqrt{q_0p_0}}\bar{u}_r\gamma^0\mathcal{E}\gamma^0 v_s=-\frac{1}{\sqrt{1+\rho^2}}
\begin{pmatrix}1&i\rho\\i\rho&1\end{pmatrix} \;.
\ee

If we consider the contribution from a single field pulse then $\epsilon=1$ and $\epsilon=-1$ give the same result,
\be
\mathcal{S}[\epsilon_{(\LCperp)}]=\frac{2\rho^2}{1+\rho^2}
\qquad
\mathcal{S}[\epsilon_{(\LCpara)}]=1 \;.
\ee
The ratio 
\be\label{polRatioOne}
\frac{P_{(\LCperp)}}{P_{(\LCpara)}}=\frac{\mathcal{S}[\epsilon_{(\LCperp)}]}{\mathcal{S}[\epsilon_{(\LCpara)}]}=\frac{2\rho^2}{1+\rho^2}
\ee
agrees with the result for a constant field in~\cite{Dunne:2009gi}.

Collecting the various contributions gives the following formula for the contribution from a single instanton,
\be
\begin{split}
\hat{M}_j=&\underbrace{2\sqrt{p_0q_0}\Sigma}_{\bar{u}\gamma^0\mathcal{E}\gamma^0v}\frac{T}{2}\underbrace{\frac{1}{(2\pi T)^{\frac{D}{2}}}}_{\eqref{pathIntRatio}\&\eqref{thetaInt}}\underbrace{\frac{\pi^D}{\sqrt{\text{det }{\bf H}}}}_{\eqref{Xint}}\underbrace{\frac{1}{\sqrt{\text{det }\Lambda}}}_{\eqref{detLambdaBW}}\underbrace{e^\varphi}_{\eqref{varphiInt}} \\
=&\sqrt{\frac{(2\pi)^D}{q_0p_0 h}}\Sigma e^\varphi \;,
\end{split}
\ee
where $h$ is given by~\eqref{hNumBW}. There is at least one instanton for each field maximum, so the total amplitude is given by a sum, which leads to interference patterns in the probability,
\be
\hat{P}=\Big|\sum_j\hat{M}_j\Big|^2 \;.
\ee
In $\hat{M}$ and $\hat{P}$, we have not included factors of $2\pi$ coming from the delta function in~\eqref{varphiIntDelta} or from $\ud^3{\bf p}/(2\pi)^3$ and $\ud^3{\bf q}/(2\pi)^3$, nor the factor of $e^2/2\Omega$. The reason is that we have anyway not included a photon wave packet here. We have already shown how to include a wave packet in~\cite{DegliEsposti:2023fbv}, but here we are mainly interested in comparing and checking the SWF formulas for BW, and in this first paper on this topic we content ourselves with plane-wave photons.

\subsection{Exact spin dependence}

The spin dependence in~\eqref{Mpara} is actually exact. To show this, we start by noting that the free spinors~\eqref{uv} can be expressed as 
\be\label{uvt}
\begin{split}
u_s({\bf p})&=\frac{(p_0+p_1+M(p_2,p_3))\tilde{R}_s}{\sqrt{2p_0(p_0+p_1)}} \\
v_s({\bf p})&=\frac{(-p_0-p_1+M(-p_2,-p_3))\tilde{R}_s}{\sqrt{2p_0(p_0+p_1)}} \;,
\end{split}
\ee
where
\be
M(p_2,p_3)=\gamma^0(1-p_2\gamma^2-p_3\gamma^3) 
\ee
and $\tilde{R}_s=\gamma^0 R_s$ satisfies the same relations as in~\eqref{Reqs}, except for the sign of $\gamma^0\gamma^1$,
\be
\gamma^0\gamma^1\tilde{R}_s=-\tilde{R}_s
\qquad
i\gamma^2\gamma^3 \tilde{R}_s=s\tilde{R}_s 
\qquad
\tilde{R}_r^\dagger \tilde{R}_s=\delta_{rs} \;.
\ee
The Dirac equation can, for $E(t,x)=-\partial_x A_0$, be expressed in terms of $M$ and $\gamma^0\gamma^1$ as
\be
-\partial_0\psi=(iA_0+iM+\gamma^0\gamma^1\partial_x)\psi \;,
\ee
but, since $\gamma^0\gamma^1 M=-M \gamma^0\gamma^1$, $\gamma^0\gamma^1\tilde{R}=-\tilde{R}$ and $M^2=1+p_2^2=m_\LCperp^2$, $M$ is the only matrix that can appear in the solution. We can therefore write
\be\label{UVM}
U_s^{\rm out}=(f_0+f_1 M[p_\LCperp])\tilde{R}_s \qquad V_s^{\rm out}=(g_0+g_1 M[-q_\LCperp])\tilde{R}_s \;,
\ee
where $f_i(t,{\bf x},p_1,p_\LCperp^2)$ and $g_i(t,{\bf x},q_1,q_\LCperp^2)$ are unknown scalar functions which do not depend on the spin.

For parallel polarization, $\epsilon_\mu=(0,1,0,0)$, the matrix element is given in terms of these scalar functions by
\be\label{me1}
\begin{split}
&U_{\rm out}^\dagger(r,{\bf p})\gamma^0\gamma^1 V_{\rm out}(s,{\bf q})=-f_0^*g_0\delta_{rs}\\
&+f_1^*g_1[(1-p_\LCperp q_\LCperp)\delta_{rs}+\delta_{r,-s}(p+q)_\LCperp\tilde{R}_r^\dagger\gamma^\LCperp\tilde{R}_s] \;.
\end{split}
\ee
With the full inner product,
\be
(\psi_1|\psi_2)=\int\ud^3{\bf x}\,\psi_1^\dagger\psi_2 \;,
\ee
we have $(U_r^{\rm out}({\bf p})|V_s^{\rm out}({\bf q}))=0$ for any values of ${\bf p}$, ${\bf q}$, $r$ and $s$. For $q_\LCperp=-p_\LCperp$ we also have 
\be\label{reducedInner}
(U_r^{\rm out}(p_1,p_\LCperp)|V_s^{\rm out}(q_1,-p_\LCperp))_1=0 \;,
\ee
where
\be
(\psi_1|\psi_2)_1=\int\ud x\,\psi_1^\dagger\psi_2 \;.
\ee
Plugging~\eqref{UVM} into~\eqref{reducedInner} gives
\be\label{f0g0f1g1}
\int\ud x(f_0^* g_0+m_\LCperp^2f_1^* g_1)=0 \;.
\ee
Since $f_i$ and $g_i$ only depend on $p_\LCperp$ and $q_\LCperp$ via $p_\LCperp^2$ and $q_\LCperp^2$, \eqref{f0g0f1g1} is valid for both $q_\LCperp=-p_\LCperp$ and $q_\LCperp=p_\LCperp$. We can therefore use~\eqref{f0g0f1g1} to simplify~\eqref{me1} for $q_\LCperp=p_\LCperp$, 
\be\label{me2}
\begin{split}
&\int\ud x\,U_{\rm out}^\dagger(r,p_1,p_\LCperp)\gamma^0\gamma^1 V_{\rm out}(s,q_1,p_\LCperp) \\
&=2[\delta_{rs}+\delta_{r,-s}p_\LCperp\tilde{R}_r^\dagger\gamma^\LCperp\tilde{R}_s]\int\ud x\,f_1^*g_1 \;.
\end{split}
\ee
With $p_3=0$ we thus find that the spin dependent part of the amplitude is given by
\be\label{MparaE}
M(\epsilon_\LCpara,r,s)=[\delta_{rs}-\delta_{r,-s}ip_2]M(\epsilon_\LCpara) \;,
\ee
so the instanton prediction~\eqref{Mpara} is in fact exact.

\end{document}